\documentclass{aa}  

\usepackage{graphicx}
\usepackage[varg]{txfonts}
\usepackage{lipsum}
\usepackage{subcaption}         
\usepackage{lscape}             
\usepackage{placeins}           
\usepackage{natbib}
\usepackage[version=4]{mhchem}
\usepackage{orcidlink}

\newcommand{\kms}{{\ensuremath{{\rm km\, s^{-1}}}}\xspace}
\newcommand{\Msun}{\ensuremath{\rm M_{\sun}}\xspace}
\newcommand{\Jyb}{\ensuremath{\rm Jy\,beam^{-1}}\xspace}
\newcommand{\vlsr}{\ensuremath{\rm v_{\mathrm{LSR}}}\xspace}
\newcommand{\SOt}{\ensuremath{\rm \ce{SO2}}\xspace}
\newcommand{\HtCO}{\ensuremath{\rm \ce{H2CO}}\xspace}

\begin{document}

   \title{PRODIGE -- envelope to disk with NOEMA}

   \subtitle{VIII. Sulfur oxides trace a shock caused by a streamer in the inner envelope of a protostar}

   \author{Mar\'ia Teresa Valdivia-Mena\inst{1,2}\orcidlink{0000-0002-0347-3837}
        \and Jaime E. Pineda\inst{2}\orcidlink{0000-0002-3972-1978}
        \and Caroline Gieser\inst{3}\orcidlink{0000-0002-8120-1765}
        \and Paola Caselli\inst{2}\orcidlink{0000-0003-1481-7911}
        \and Dominique M. Segura-Cox\inst{4}\orcidlink{0000-0003-3172-6763}
        \and Yuxin Lin \inst{2}\orcidlink{0000-0001-9299-5479}
        \and Mar\'ia Jos\'e Maureira \inst{2}
        \and Tien-Hao Hsieh \inst{5}
        \and Laura A. Busch\inst{2}
        \and Ana Lopez-Sepulcre \inst{6,7}
        \and Laure Bouscasse \inst{6}
        \and Dmitry Semenov \inst{8,3}
        \and Asunci\'on Fuente \inst{9}
        \and Nichol Cunningham \inst{10}
        \and Thomas Henning \inst{2}
        \and Juli\'an J. Miranzo-Pastor \inst{9}
        \and Yu-Ru Chou\inst{2}
        \and Roberto Neri \inst{6}
        \and Izaskun Jimenez-Serra \inst{9}
        \and Edwige Chapillon \inst{6}
        \and Stephane Guilloteau \inst{11}
        \and Felipe Alves \inst{6}
        \and Mario Tafalla \inst{12}
        \and Anne Dutrey \inst{11}
        \and Riccardo Franceschi \inst{13}
        \and Sierk van Terwisga \inst{14}
        \and Kamber Schwarz \inst{3}
        }

   \institute{European Southern Observatory, Karl-Schwarzschild-Strasse 2 85748 Garching bei Munchen, Munchen, Germany 
   \email{mariateresa.valdiviamena@eso.org}
   \and
   Max-Planck-Institut f\"{u}r extraterrestrische Physik, Giessenbachstrasse 1, D-85748 Garching, Germany 
   \and
   Max-Planck-Institut f\"ur Astronomie, K\"onigstuhl 17, 69117 Heidelberg, Germany 
   \and
   Department of Physics and Astronomy, University of Rochester, Rochester, NY 14627-0171, USA
   \and
   Taiwan Astronomical Research Alliance (TARA), Taiwan; Institute of Astronomy and Astrophysics, Academia Sinica, PO Box 23-141, Taipei, 106, Taiwan 
   \and
   Institut de Radioastronomie Millim\'{e}trique (IRAM), 300 rue de la Piscine, F-38406, Saint-Martin d'H\`{e}res, France 
   \and
   IPAG, Universit\'{e} Grenoble Alpes, CNRS, F-38000 Grenoble, France 
   \and
   Zentrum f\"{u}r Astronomie der Universit\"{a}t Heidelberg, Institut f\"{u}r Theoretische Astrophysik, Albert-Ueberle-Str. 2, 69120 Heidelberg, Germany
   \and
   Centro de Astrobiolog\'{\i}a (CAB), CSIC-INTA, Ctra. de Torrej\'on a Ajalvir km 4, 28806, Torrej\'on de Ardoz, Spain 
   \and
   SKA Observatory, Jodrell Bank, Lower Withington, Macclesfield SK11 9FT, United Kingdom
   \and 
   Laboratoire d'Astrophysique de Bordeaux, Universit\'{e} de Bordeaux, CNRS, B18N, All\'{e}e Geoffroy Saint-Hilaire, F-33615 Pessac, France
   \and 
   Observatorio Astron\'{o}mico Nacional (IGN), Alfonso XII 3, E-28014, Madrid, Spain
   \and 
   Universit\'{e} Paris-Saclay, CNRS, Institut d'Astrophysique Spatiale, 91405, Orsay, France
   \and 
   Space Research Institute, Austrian Academy of Sciences, Schmiedlstr. 6, 8042, Graz, Austria
   }

   \date{Received November 4, 2025, Accepted March 14, 2026}

 
  \abstract
   {Recently, streamers have been observed causing shocks at the outer edge of protoplanetary disks. The study of sulfur-bearing species can help us to understand the physical and chemical changes caused by infalling streamers toward their landing positions.}
   {We study the physical properties traced by the emission of \SOt and SO toward the Class I protostar Per-emb 50, which is possibly related to the streamer infalling toward its disk.} 
   {We present new NOrthern Extended Millimeter Array (NOEMA) A-array observations as part of the large program "Protostars and Disks: Global Evolution" (PRODIGE). We analyzed the morphology of \SOt and SO, and complement our interpretations with additional \HtCO and CO data from the same program. We compared the \SOt and SO morphology with an infalling-rotating model. We applied Bayesian model selection to the brightest \SOt line to disentangle the different kinematic components traced by this molecule. We used Local Thermodynamic Equilibrium (LTE) and non-LTE analyses to determine the temperature and density of the \SOt emission. }
   {There are two separate peaks of \SOt emission offset toward the southwest of Per-emb 50, one brighter (peak 1) at about 180 au from the protostar, and a weaker one (peak 2) at about 400 au. Peak 2 is blueshifted with respect to an infalling-rotating envelope. We propose that this peak is caused by the shock between the inner envelope and the streamer. Peak 1 is consistent with the expected envelope motion, and could thus be caused by shocks at the disk-envelope interface, but potential streamer influence cannot be neglected. Both peaks show abundance ratios consistent with a low velocity shock ($\sim 3-4$ \kms) when compared with shock models.}
   {Streamers can affect the physical and chemical structure of both disks and envelopes, suggesting that streamers can play an important role in shaping both structures in the embedded stages of star formation.}


   \maketitle

\nolinenumbers 
\section{Introduction}

The evolution of protostars and their planet-forming disks is fundamental to understanding how planets (such as Earth) are formed. Protostars are part of a complex, dynamic system, with energetic outflows ejecting material and transporting angular momentum, and infalling material feeding the disk simultaneously \citep[][and references within]{Tobin2024revprotostars}. These processes can have a deep impact in the physical \citep[e.g.][]{Hennebelle2017disksimspiralarms, Zamponi2021hotdisk, Kuznetsova2022asyminfall, Maureira2022dusthotspots} and chemical properties of disks \citep[e.g.][]{Sakai2014NatcentbarrierSO, Vastel2022methanolshocks, Podio2024streamer, Hsieh2025ch3cnpaper}. In particular, the chemical properties of protostellar disks are fundamental to understand what materials are inherited from the interstellar medium (ISM), and what has been reprocessed within the disk \citep{Pontoppidan2014volatilesreview, Drozdovskaya2018sulfurincomets, Ceccarelli2023organichchemrev}.

Infall of envelope material causes a shock when entering the disk. This shock heats up the surrounding gas, potentially increasing the temperature to hundreds of K \citep{Draine1983shocks, Neufeld1994shocks}. Thus, shocks can change both the conditions of the material fed to the disk and within the disk itself. Given the hints that planet formation begins in these early, embedded stages \citep{Manara2018diskmassesexopl,Tychoniec2020dustmassClassI, Sheehan2020substructureClassI, Segura-Cox2020IRS63, Hsieh2025campos}, investigating how disk and envelope material changes is essential to understanding our chemical inheritance.

Not all infall from the environment comes from a symmetric envelope. Asymmetries in envelopes are common toward low-mass protostars \citep[e.g.][]{Tobin2012asymmetries}. Recently, streamers have been observed toward a variety of protostars, of different masses, ages, and located across different star-forming regions \citep[][and references within]{Pineda2023PPVII, Tobin2024revprotostars}. These can bring material from beyond the envelope \citep{Pineda2020NatAsPer2, Valdivia-Mena2023B5, Taniguchi2024Per2, Gieser2025antiheroes},  expanding the material reservoir available for stellar and planetary growth. In some cases, the streamer's chemical footprint suggest the funneled mass is prestellar in nature \citep{Taniguchi2024Per2, Podio2024streamer}. The study of the effects of streamers onto protostellar \citep[e.g.][]{Flores2023eDisk,Artur2022IRS44shocks,Aso2023eDisk,Podio2024streamer, Liu2025SOSO2streamers, Kido2025eDisk} and protoplanetary disks \citep{Garufi2022almadotshocks,Speedie2025ABAurshock} has been an active area of research in the last few years. It is uncertain if the material brought by streamers can be directly inherited by forming planets or if its chemically reset due to the impact between them and disks. Therefore, understanding the shock conditions produced by streamers is fundamental to determining the chemical effects of streamers. 

Sulfur-bearing molecules are commonly used as tracers of heated regions, such as shock fronts, given their sensitivity to the physical conditions of their environment \citep{Sakai2014NatcentbarrierSO, Oya2025SOinjetElias29}. In particular, SO and \SOt have been associated with shocks caused by both envelope infall \citep{Sakai2014NatcentbarrierSO} and streamers \citep{Garufi2022almadotshocks, Speedie2025ABAurshock}, sometimes with both phenomena occurring at the same time \citep{Liu2025SOSO2streamers}. SO and \SOt emission are sensitive to shock conditions (such as velocity and pre-shock number density $n_{\mathrm{H}}$) as its formation pathways in the gas phase are closely tied to shock-related processes \citep[e.g.,][]{vanGelder2021shockmodel}. Both species can be sublimated from dust grains in shocks and also formed in the gas phase due to the temperature increase caused by shocks, but SO thermal desorption requires lower velocities and densities than \SOt as its sublimation temperature is lower \citep[37 K for SO versus 62 K for \SOt,][]{vanGelder2021shockmodel}, and thus the former can be more easily detected \citep{Aota2015SOshock}. The study of these sulfur-bearing species can then help us understand the conditions of shocks caused by streamers.

In this article, we investigate the emission of the sulfur-bearing molecules \SOt and SO toward Per-emb 50, a Class I protostar with a known streamer. Per-emb 50 is located in NGC~1333, an active star-forming region within the Perseus Molecular Cloud, at a distance of 298 pc \citep{Zucker2018dist}. Previously, \cite{Valdivia-Mena2022prodigeI} found a streamer toward this source in \HtCO $3_{0,3}-2_{0,2}$ emission, with an extension of at least 3300 au in the southwest direction, transporting material from beyond the envelope toward the vicinity of the disk. In the same work, they found an unresolved \SOt emission peak offset from the protostar's position toward the southwest, and suggested this was caused by the accretion shock from the streamer material impacting near the disk. Later, data from the Perseus ALMA Chemistry Survey \citep[PEACHES,][]{Zhang2023PEACHES} showed that the enhancement of SO and \SOt toward the southwest coincided with a change in the dust properties of the envelope, consistent with an accretion shock. We have obtained new NOEMA observations using the most extended configuration (A), which allows us to investigate the \SOt and SO emission in more detail and answer if the sulfur species' emission is consistent with a shock caused by the streamer.

This paper is organized as follows. Section \ref{sec:observations} presents the new NOEMA data, which is part of the large program ``Protostars and Disks: Global Evolution'' (PRODIGE). 
This program has been fruitful for the discovery of streamers toward low-mass protostars \citep{Valdivia-Mena2022prodigeI, Hsieh2023SVS13streamer, Gieser2024AL1448N, Gieser2025antiheroes}. In Sect. \ref{sec:results}, we show the physical properties found through \SOt  and SO emission and explains the steps we took to arrive to these results. In Sect. \ref{sec:discussion}, we argue that part of the observed \SOt and SO emission is caused by the streamer producing a shock in the inner envelope of the protostar. A summary of this work is presented in Sect. \ref{sec:summary}.


\section{Observations and data reduction \label{sec:observations}}

We used the NOEMA C and D array observations for Per-emb 50 from the PRODIGE Max Planck - IRAM Observing program. Per-emb 50 was observed with C configuration on 29 December 2019 and 05 January 2020. Details of the data reduction are in \cite{Valdivia-Mena2022prodigeI}. In summary, we used the data reduction pipeline available in \texttt{CLIC} within the IRAM servers, and combined the uv tables using \texttt{uvmerge}. The C and D combined data were phase self calibrated with solution intervals of 30, 135 and 45 s, using the \texttt{mapping} self-calibration module, and then applied the phase solutions to the continuum and line data.

Observations with NOEMA A configuration were taken during 19 and 21 January 2023 under project W22AH (PI: Valdivia Mena). 
We designed the A observations with the same spectral setup as the PRODIGE observations so that the data can be directly combined. We calibrated the data using the NOEMA pipeline available through the \texttt{GILDAS} \texttt{CLIC} package. 
We self-calibrated the data using the \texttt{GILDAS mapping} self-calibration module. The iterative self-calibration was performed with solution intervals of 300 s, 135 s, and 66 s on the continuum data, which includes only line-free channels. We applied the phase gain solutions to the continuum and line uv tables. 

We combined and imaged the data using the \texttt{MAPPING} package from the \texttt{GILDAS} suite\footnote{https://www.iram.fr/IRAMFR/GILDAS}. 
We used the self-calibrated uv tables from the PRODIGE data reduction process and the self-calibrated A-configuration uv tables. We combined the A and CD array uv tables using \texttt{uv\_merge}. 
Continuum subtraction was done manually on the combined uv tables using the \texttt{uv\_baseline} command.

We imaged the SO, \SOt, CO, and \HtCO line cubes from the combined arrays' narrowband units, which have a native resolution of 62.5 kHz (approximately 0.08 \kms for the rest frequencies of the investigated lines). We imaged using the classical Hogbom CLEAN algorithm \citep{Hogbom1974clean} with natural weighting. We manually masked the emission in the deconvolved image to improve the image quality and reduce emission sidelobes. The data cubes have a spatial resolution of approximately $0.8\arcsec \times 0.3\arcsec$, with a position angle (PA) of $-0.2$ deg. Their properties are summarized in Table \ref{tab:cubeprops}.

We smoothed all \SOt data cubes to a resolution of 0.21 \kms to increase the signal-to-noise ratio ($SNR$) of \SOt transitions $4_{2,2}-3_{1,3}$ and $12_{3,9}-12_{2,10}$, which are weaker than \SOt $11_{1,11}-10_{0,10}$. The cubes after smoothing have an rms of about 0.5 K (Table \ref{tab:cubeprops}).  

We also obtained the continuum emission toward Per-emb 50 from the wideband units. We imaged the continuum using the Hogbom CLEAN algorithm with robust weight (Briggs), with a robust parameter value of 1, and applied manual masking to improve image quality. The continuum is unresolved for the resolution of our data ($0.78 \times 0.28 \arcsec$, $-180.26^{\circ}$) and has an rms of 0.15 m\Jyb. The location of the protostar used throughout this work is the location of the maximum peak in the continuum image (RA(J2000) $=3^h09^m07^s.77$, DEC(J2000) $= +31^{\circ}21\arcmin56\farcs99$).

\begin{table}[h]
\caption{\label{tab:cubeprops}Molecular transitions and their properties}
\centering
\begin{tabular}{lllll}
\hline\hline
Molecule& Transition&$\nu$\tablefootmark{a} &$E_{\mathrm{up}}$ &rms\\
&&(GHz)&(K) &(K)\\
\hline
SO & $5_{5} - 4_{4}$ & 215.221 & 44.10 & 0.85 \\
\ce{H2CO} & $3_{0,3} - 2_{0,2}$ & 218.222 & 20.96 & 0.69 \\
\ce{SO2} & $11_{1,11}-10_{0,10}$ & 221.965 & 60.36 & 0.44\tablefootmark{b}\\
  & $4_{2,2}-3_{1,3}$  &  235.152 & 19.03 & 0.45\tablefootmark{b} \\
  & $12_{3,9}-12_{2,10}$ & 237.069  &  93.96 & 0.50\tablefootmark{b} \\
CO & $2-1$ & 230.538 & 16.60 & 0.61 \\
\hline
\end{tabular}
\tablefoot{
\tablefoottext{a}{Frequencies and upper energies obtained from the CDMS catalog \citep{Endres2016CDMS}.}
\tablefoottext{b}{Noise level after smoothing the spectral axis to 0.21 \kms.}
}
\end{table}


\section{Results and analysis\label{sec:results}}

\subsection{Morphology of molecular emission\label{sec:morphology}}

\begin{figure*}
    \centering
    \includegraphics[width=0.98\linewidth]{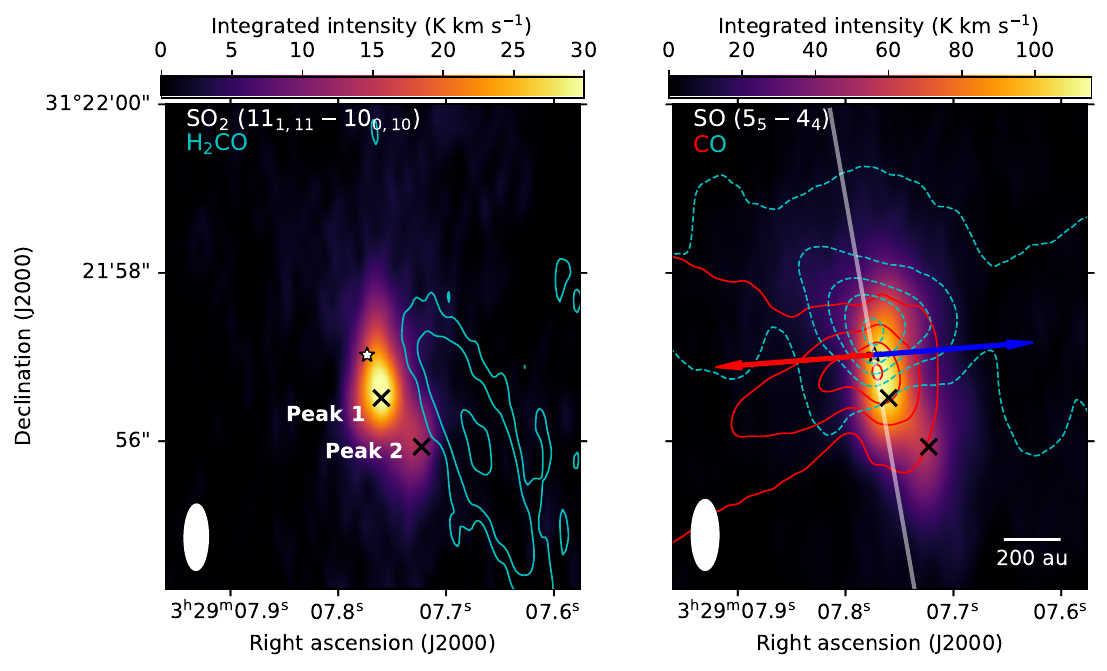}
    \caption{NOEMA observations of molecular emission toward Per-emb 50. Left: Integrated intensity of \SOt $11_{1,11}-10_{0,10}$ between 6.5 and 12 \kms. Black crosses show the locations of the \SOt peaks, labeled as peak 1 and peak 2. Cyan contours represent the \HtCO $3_{0,3} - 2_{0,2}$ integrated intensity between 5 and 8 \kms, drawn in steps of 3, 5 and 10 times rms of the integrated image (0.7 K \kms). Right: Integreated intensity of SO $5_{5} - 4_{4}$ between 0 and 13 \kms. Red and cyan contours are CO integrated intensity in redshifted and blueshifted channels, respectively, with respect to the protostar's \vlsr (7.5 \kms). Blueshifted channels are integrated between -4.3 and 5.3 \kms, whereas redshifted channels are between 10 and 20 \kms. Contours are drawn at 5 to 45 times the rms of the integrated images (4.7 K \kms) in steps of 10. Red and blue arrows indicate the direction of the outflow. The dashed line represents the direction of the PV diagram in Fig. \ref{fig:SO2pvdiag}. The white star marks the position of the continuum peak. The filled white ellipse represents the beam size. A scalebar in the bottom right corner represents a physical scale of 200 au.}
    \label{fig:mom0wh2co}
\end{figure*}

Figure \ref{fig:mom0wh2co} shows the integrated intensity maps of \SOt and \HtCO, together with SO and CO contours. 
The \SOt peak seen in \cite{Valdivia-Mena2022prodigeI} is resolved into two separate intensity peaks. They are seen in all \SOt transitions, but most prominently in the brightest \SOt line ($11_{1,11}-10_{0,10}$, shown in Fig. \ref{fig:mom0wh2co}). Maps of the rest of \SOt transitions are presented in Appendix \ref{ap:zoomSO2}. The location of the two \SOt peaks are offset from the position of the continuum emission in the southwest direction. The brightest peak, which we label peak 1, is located $\sim180$ au (0.6\arcsec) from the protostar, whereas the second, weaker peak (peak 2) is $\sim400$ au (1.3\arcsec) away. There is also weak \SOt emission toward the north of the protostar, seen only in the brightest transition. 

The brightness distribution of SO is more extended than \SOt, with strong emission ($SNR>5$) toward both north and south of the protostar, but the brightest emission both in integrated intensity and in peak brightness temperature ($T_{\mathrm{MB}}$) coincide with \SOt peaks 1 and 2 (Fig. \ref{fig:mom0wh2co} right). \cite{Valdivia-Mena2022prodigeI} showed that SO emission can be fit with up to three velocity components that trace different structures, included the streamer toward the southwest, but there was only one beam-sized emission peak south of the protostar in the C and D configuration data (beam of 1.2\arcsec). Adding the A configuration, this peak resolves into two distinct emission peaks (at a resolution of 0.8\arcsec).

\HtCO distribution peaks toward the west of the \SOt peak 2 and is extended, reaching approximately the same declination as the protostar. This new data resolves the part of the streamer first seen in \cite{Valdivia-Mena2022prodigeI} that shows the strongest velocity change, accelerating toward blueshifted velocities with respect to the protostar. There is no \SOt emission at the location of \HtCO and vice-versa (Appendix \ref{ap:zoomSO2}), but there is SO emission at the location of the streamer with the same velocity. 

CO traces the outflow cones that are approximately in the east-west direction, as shown in the contours of Fig. \ref{fig:mom0wh2co} right. At this resolution, CO emission is also tracing part of the inner envelope and disk surrounding the protostar, hinted by the peaks of both redshifted and blueshifted CO located south and north of the protostar, as shown in Appendix \ref{ap:zoomSO2}, aligned with the previously known rotation from the envelope and disk \citep{Valdivia-Mena2022prodigeI, Zhang2023PEACHES}. Based on the integrated intensity maps, \SOt peak 1 appear close to the base of the outflow, where CO shows redshifted emission. However, peak 2, farther away, is not associated with high-velocity CO emission.

\subsection{Kinematics traced by SO$_2$ and SO} 

\begin{figure*}[h]
    \centering
    \includegraphics[width=0.95\textwidth]{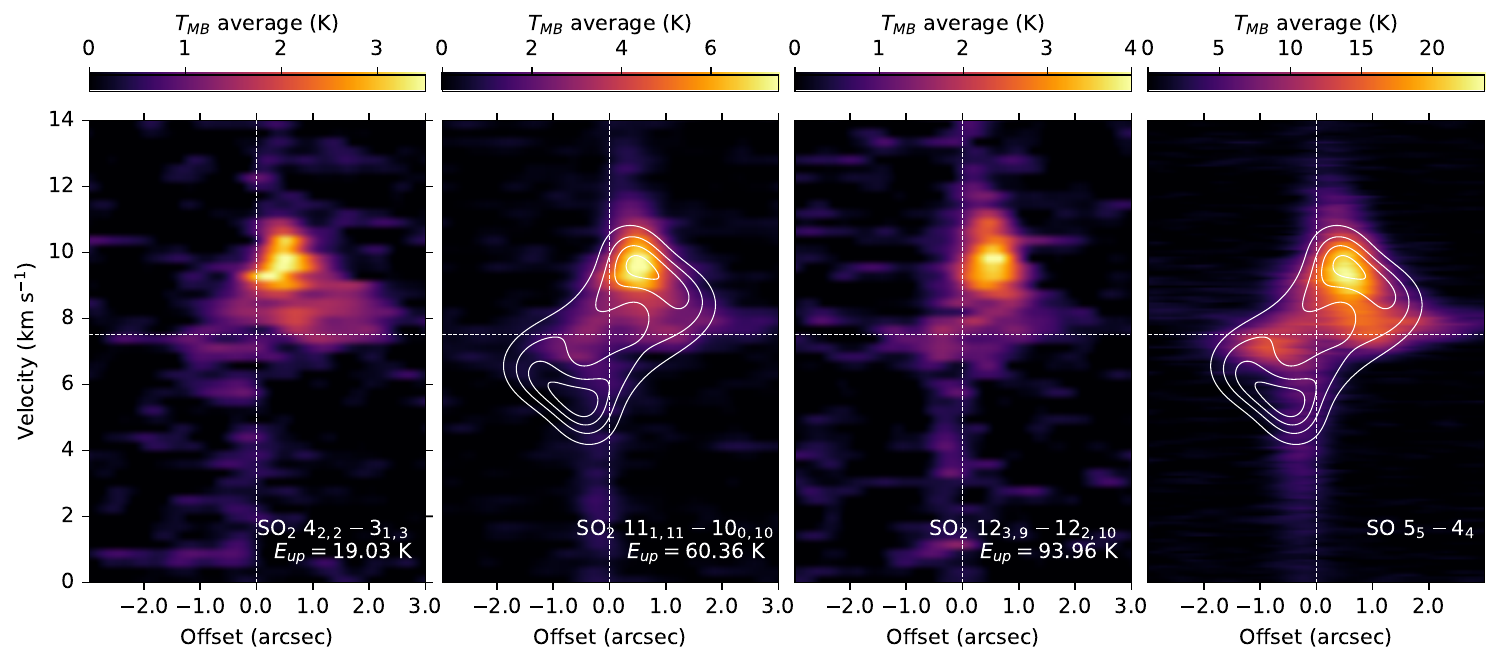}
    \caption{Position velocity diagrams of \SOt and SO using the path shown in Fig. \ref{fig:mom0wh2co}. \SOt transitions are in order of increasing upper energy levels $E_{\mathrm{up}}$ From left to right: \SOt $4_{2,2}-3_{1,3}$, \SOt $11_{1,11}-10_{0,10}$, \SOt  $12_{3,9}-12_{2,10}$, SO $5_{5} - 4_{4}$. White contours show the normalized intensity PV diagrams for an infalling-rotating envelope, obtained with FERIA \citep{Oya2022FERIA}.}
    \label{fig:SO2pvdiag}
\end{figure*}

\begin{figure}[]
    \centering
    \includegraphics[width=0.95\linewidth]{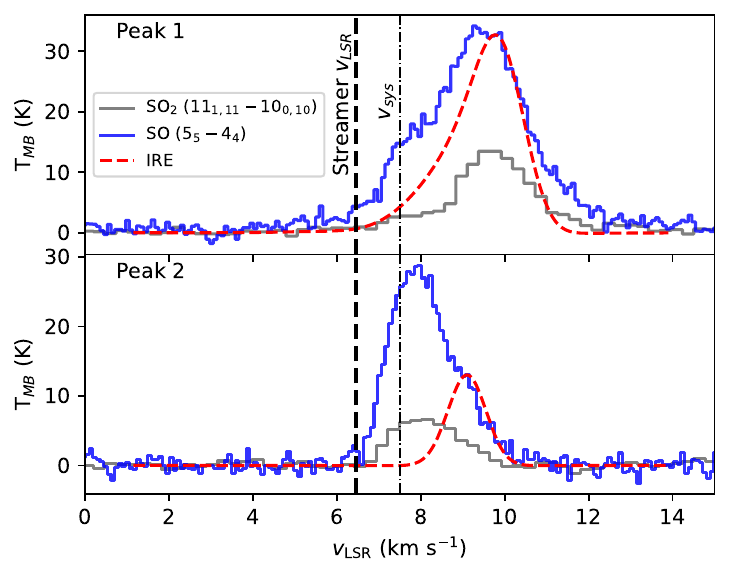}
    \caption{Spectra of SO $5_{5} - 4_{4}$ (blue) and \SOt $11_{1,11}-10_{0,10}$ (gray) at the two resolved peak positions, together with the spectra obtained from FERIA, normalized to the SO intensity at the peak velocity of the IRE. The vertical dashed-dotted line marks the systemic velocity of the protostar (7.5 \kms), whereas the thick dashed line marks the velocity of the streamer at the same distance from the protostar (in radius) as peak 2 (6.45 \kms, Appendix \ref{ap:streamerH2CO}).  }
    \label{fig:IREspectra}
\end{figure}

We made position-velocity (PV) diagrams for our \SOt and SO emission cubes along the white line shown in Fig. \ref{fig:mom0wh2co} right (190 deg from North), which corresponds to the envelope position angle found in \cite{Zhang2023PEACHES}, to check if our \SOt and SO detected emission traces rotational signatures. We compared the emission in both peaks with an infalling-rotating envelope (IRE) model obtained using the FERIA code \citep[Flat Envelope Model with Rotation and Infall under Angular Momentum Conservation,][]{Oya2022FERIA} to further analyze the kinematics of peaks 1 and 2. We describe how we obtained the IRE model PV diagram in Appendix \ref{ap:zoomSO2}. The resulting PV diagrams are in Fig. \ref{fig:SO2pvdiag}.

Both \SOt and SO transitions show a clear redshifted peak at approximately 9.5 \kms in the PV diagrams, which corresponds to peak 1 in the moment 0 map of Fig. \ref{fig:mom0wh2co}. This peak is located at the expected position and velocity for the redshifted emission peak of the IRE model described in \cite{Zhang2023PEACHES}. The rest of the emission in that quadrant has lower velocities than those expected from the IRE model. This brightness distribution is associated to peak 2 from the \SOt moment 0 map. There is both redshifted and blueshifted emission within a beam of the protostar with high velocity (up to 5 \kms from the \vlsr), probably associated with a Keplerian disk around the protostar, seen in all transitions except for \SOt $12_{3,9}-12_{2,10}$, the one with highest $E_{\mathrm{up}}$ (93 K). Both molecules show a skewed diamond shape which is an indication of a combination of rotation and infall gas motions.

Figure \ref{fig:IREspectra} shows the \SOt and SO spectra at the two peak positions (crosses in Fig. \ref{fig:mom0wh2co}) together with the expected line shape for emission corresponding to an IRE. The expected IRE velocity in the envelope at peak 1 coincides with the observed SO and \SOt maxima, whereas for the peak 2, \SOt and SO emission is blueshifted from the expected IRE peak (9.1 \kms) by approximately 1.4 \kms. Both molecules' emission could also include a component of IRE, seen as a redshifted shoulder in the spectra, but the peak of the emission does not correspond to envelope kinematics. For comparison, we label the velocity of the streamer head obtained through the fit of \HtCO velocity components (6.45 \kms, Appendix \ref{ap:streamerH2CO}). Peak 2 velocity is in between the streamer velocity and the expected IRE velocity at that location.

\subsection{Separation of individual velocity components \label{sec:nestedsampling}}

The PV diagrams in Fig. \ref{fig:SO2pvdiag} and spectra in Fig. \ref{fig:IREspectra} suggest that there is more than one kinematic component traced by S-bearing molecules. To investigate the physical nature of the \SOt peaks, we fit up to three Gaussian velocity components to the \SOt $11_{1,11}-10_{0,10}$ spectra in each pixel. We applied Bayesian model selection with the Nested sampling parameter exploration algorithm, using the \texttt{pyspecnest} Python package \citep{Sokolov2020BayesFit}. Details of the application of this method to molecular line observations are in \cite{Sokolov2020BayesFit}. In summary, Nested Sampling algorithm searches for the best parameters for each model, where in this case, the different models are one, two, and three Gaussian components, and returns the Bayes factor $K$ for each pair of models, which we use to select the optimal number of Gaussian components in each pixel. We applied the algorithm to all spectra with $SNR>5$, and each pixel is assumed independent from each other. Details of this process are in Appendix \ref{ap:gaussfits}.

\begin{figure}
    \centering
    \includegraphics[width=0.99\linewidth]{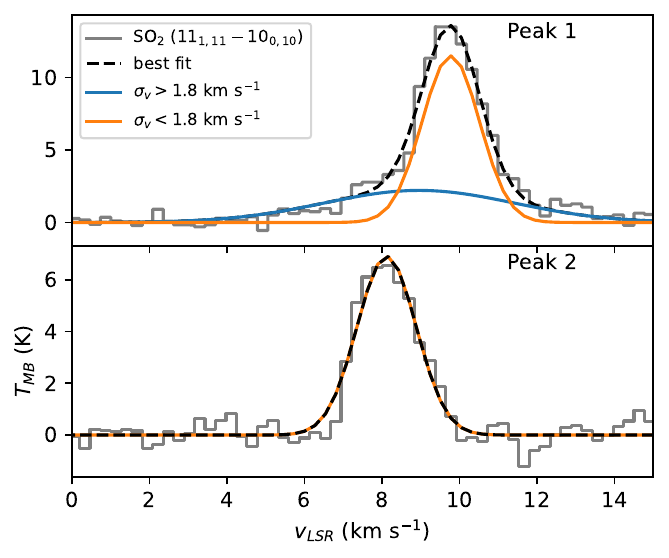}
    \caption{\SOt spectra (gray) at the location of the peaks together with the best fit (black dashed lines) Gaussian models. Peak 1 has two Gaussian components, a narrow (orange, $\sigma_v<1.8$ \kms) and a wide (blue, $\sigma_v>1.8$ \kms), whereas peak 2 has just one component (narrow).}
    \label{fig:SO2specwithfit}
\end{figure}

Figure \ref{fig:SO2specwithfit} shows the best Gaussian fits at the locations of the two \SOt peaks. Peak 1 is best fit using two Gaussian components, a wide ($\sigma_v \sim 4$ \kms) and a narrow one ($\sigma_v =0.7$ \kms), whereas peak 2 is best fit by one narrow Gaussian component, according to our model selection criterion (Appendix \ref{ap:gaussfits}). The Gaussian fit does not capture the skewness shown by the peak 2's line profile, better seen in Fig. \ref{fig:IREspectra}, but this does not affect our subsequent analysis. The Gaussian components that dominate the flux in both locations have similar velocity dispersions (peak 1 has $\sigma_v =0.7$ \kms, and peak 2, $\sigma_v =0.8$ \kms).  

\begin{figure}
    \centering
    \includegraphics[width=0.99\linewidth]{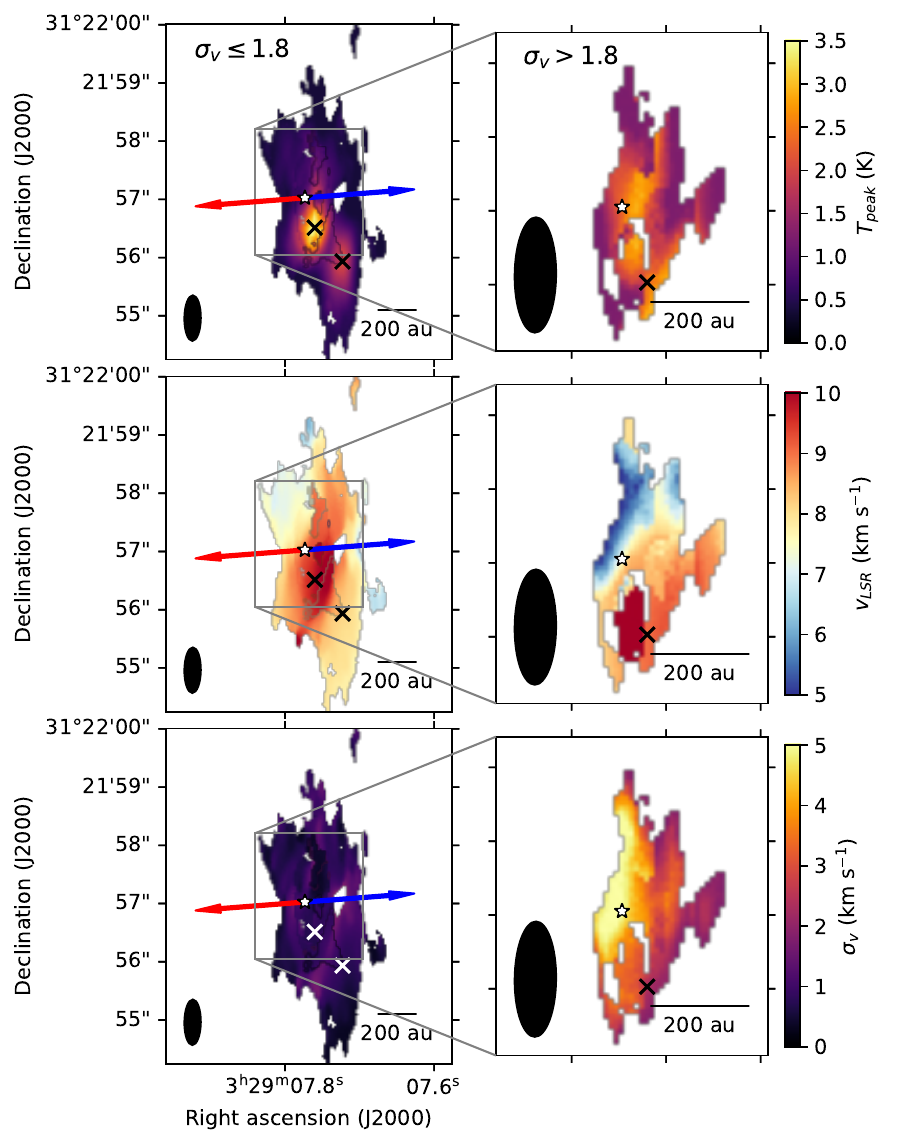}
    \caption{Peak main beam temperatures (top), central velocities (middle) and velocity dispersions (bottom) of the \SOt Gaussian components, separated into $\sigma_v\leq1.8$ \kms (left) and $\sigma_v>1.8$ \kms (right). The images in the right column focus on a region approximately 250 au in radius from the protostar. Blue and red arrows mark the direction of the outflow. Black ellipses represent the beam size. Crosses (black and white) represent the locations of the \SOt peaks. A scalebar indicates a length of 200 au.} 
    \label{fig:bayessigmasep}
\end{figure}

The fit results indicate that close to the protostar, one of the Gaussian components has a large $\sigma_v$ value with respect to the majority of pixels. We determined the $\sigma_v$ threshold to separate between narrower and wider Gaussian components close to the protostar by estimating the probability density function (PDF) of the $\sigma_v$ distribution. Details of this estimation and the resulting distribution of $\sigma_v$ values is in Appendix \ref{ap:gaussfits} (Fig. \ref{fig:sigmasepSO2}). The majority of the Gaussian components have $\sigma_v\approx 1$ \kms and there is an inflection point around $\sigma_v=1.8$ \kms. We separated the Gaussian component with $\sigma_v>1.8$ \kms (wide) from the rest of the components (narrow) to analyze both separately. In the locations where there are two components with $\sigma_v<1.8$, we chose the one which had the velocity closest to its neighbors. The velocity patterns when separating both components by $\sigma_v$ are in Fig. \ref{fig:bayessigmasep}.

The two Gaussian components (narrow and wide) show patterns consistent with rotation, with the northeast side blueshifted with respect to the protostar and the southwest redshifted.
Components with $\sigma_v>1.8$ \kms are concentrated around the protostar, in a radius of about 200 au. Their velocity pattern and large velocity dispersion suggest this \SOt component traces a rotating gas disk, consistent with the shape of the PV diagram at that location (Fig. \ref{fig:SO2pvdiag}). The narrow Gaussian components are more extended in space and include the bright peaks seen in the integrated maps (Fig. \ref{fig:mom0wh2co}). These also show a redshifted-blueshifted pattern aligned with the rotation seen in the wide component, consistent with the kinematics of the IRE seen in the PV diagram. Thus, we associate the narrow \SOt component with infalling-rotating envelope emission.

The region in between the two \SOt peaks has two narrow Gaussian components according to the Nested Sampling results (Appendix \ref{ap:gaussfits}). The fit in this region shows a potential bridge in the velocities. However, this region is smaller than one beam size, so we do not analyze this bridge further. 
We selected in this region the brightest component to show in Fig. \ref{fig:bayessigmasep}. 

\subsection{Physical conditions of the \SOt peaks\label{sec:so2lte-nonlte}}

We estimated the physical conditions traced by all three \SOt transitions (Table \ref{tab:cubeprops}) for the narrow component found in Sect. \ref{sec:nestedsampling}. Separating the disk (wide) from the envelope (narrow component) allows us to determine the densities and temperatures of each peak, avoiding contamination from the disk component. We first assumed Local Thermodynamic Equilibrium (LTE) conditions, and then used the line ratios at the location of the peaks to corroborate the obtained column densities without assuming LTE. 

We fit the three observed \SOt lines with the eXtended CASA Line Analysis Software Suite \citep[XCLASS,][]{Moller2017XCLASS}\footnote{https://xclass-pip.astro.uni-koeln.de/}, using the \texttt{myXCLASSfit} module in Python. XCLASS fits the spectra of all transitions simultaneously, assuming LTE but not optically thin emission, instead correcting for optical depth effects. We did a pixel-by-pixel LTE fit fixing the number of components, their central velocities and velocity dispersions to the values obtained from the Nested Sampling fit (Sect. \ref{sec:nestedsampling}), assumed a beam filling fraction of 1, and left the rotational temperature $T_{\mathrm{rot}}$ and the column density $N(\mathrm{SO}_2)$ as free parameters. We only fitted pixels where all three transitions show emission with $SNR>5$.

\begin{figure}
    \centering
    \includegraphics[width=0.5\textwidth]{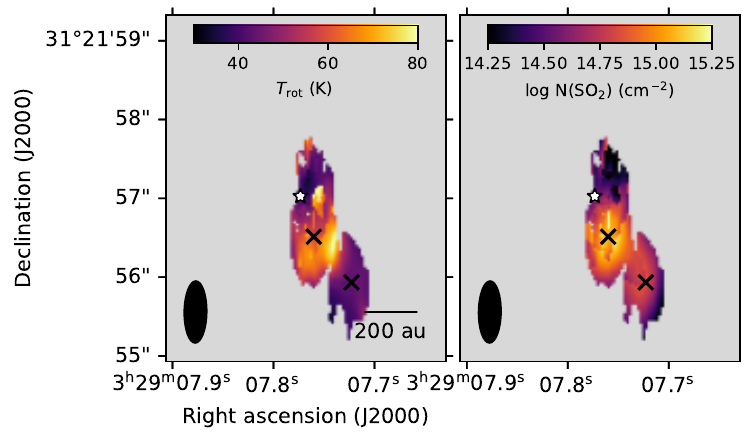}
    \caption{Parameters derived from LTE calculations of all \SOt transitions. Peak positions are marked as in Fig \ref{fig:mom0wh2co}. The black ellipse represents the beam size. A scalebar represents a length of 200 au. Crosses represent the locations of the \SOt peaks. Left: Rotational temperature. Right: Base 10 logarithm of the \SOt column density.  }
    \label{fig:XCLASSSO2narrow}
\end{figure}

The resulting temperature and column density for the narrow \SOt components are shown in Fig. \ref{fig:XCLASSSO2narrow}. The specific values obtained from the LTE fit at the location of each peak are in Table \ref{tab:LTEprops}. For the two \SOt peaks, we obtained the uncertainties in temperature and column density by running Markov Chain Monte Carlo (MCMC) algorithm, implemented in the myXCLASSfit module, whereas for the central velocities and FWHM, the uncertainties come from the Nested Sampling. Both peaks show different physical conditions. The \SOt column density at the peak 1 is about 0.2 dex higher than the column density at peak 2 (which is closer to the streamer), and the temperature of the former is about 20 K higher. The velocity dispersion, however, is similar in both locations (also seen in Fig. \ref{fig:SO2specwithfit}). The column densities are close to $10^{15}$ cm$^{-2}$, within the range estimated by \cite{Artur2023PEACHES} but higher by almost two orders of magnitude than the column densities found in \cite{Zhang2023PEACHES}. In both cases referenced above, the column density was calculated using a non-LTE calculation and only one transition of \SOt was used ($14_{0,14}-13_{1,13}$), thus introducing a high uncertainty. Nevertheless, values around $10^{15}$ cm$^{-2}$ have been reported at distances about 500 au from a protostar, such as in Elias 29 ridge region \citep{Oya2019shocksElias29} which is associated with a shocked region due to outflow interaction \citep{Oya2025SOinjetElias29}.

\begin{table}[h!]
\caption{\label{tab:LTEprops}Physical properties of the \SOt peaks from Nested Sampling and LTE analysis}
\centering
\begin{tabular}{lll}
\hline\hline
& Peak 1 & Peak 2 \\
\hline
$T_{\mathrm{rot}}$ (K) & $63.9^{+12.8}_{-16.5}$  & $46.2^{+6.8}_{-8.7}$\\
log $N(\mathrm{SO}_2)$ (cm$^{-2}$) &  $15.11^{+0.07}_{-0.08}$  & $14.84^{+0.04}_{-0.04}$ \\
$\Delta v$ (\kms)  & $1.69\pm0.06$    &$1.87\pm0.07$\\
$v_{\mathrm{LSR}}$ (\kms)  & $9.67\pm0.02$ & $8.12\pm0.03$ \\
\hline
\end{tabular}
\end{table}

Non-LTE conditions were evaluated on the basis of the line ratios of the three \SOt lines, using the RADEX radiative transfer code \citep[Appendix \ref{ap:nonLTE};][]{vanderTak200RADEX}. We ran a grid of kinetic temperatures and column densities using fixed volume densities $n_{\mathrm{H}_2}$ of $10^6$, $10^7$, and $10^8$ cm$^{-3}$. The line intensities of \SOt transitions cannot be replicated using RADEX assuming a beam filling factor of 1, so we instead compared the line ratios with the RADEX models to avoid assuming a filling factor. In particular, we compared the line ratios for peak 2 with the line ratios resulting from the different grids. Our RADEX results indicate that the density must be at least $10^7$ cm$^{-3}$ or above to explain the line ratios found in peak 2. This is close (but larger) to the critical density of the brightest \SOt line ($\approx 9\times 10^{6}$ cm$^{-3}$). 
Our results also indicate that in the $10^7-10^8$ cm$^{-3}$ regime, the line with the highest $E_{\mathrm{up}}$ (93 K) is sub-thermally excited, with $T_{\mathrm{ex}}\approx35$ K and $T_{\mathrm{kin}}\approx 45-50$ K. 
The line ratio analysis suggests that the column density $N$(\SOt) is higher than the one determined with LTE assumptions. Based on the $10^7$ cm$^{-3}$ plot (Fig. \ref{fig:radexresults} middle), log $N(\mathrm{SO}_2)$ should be closer to 15.4. This difference in column density between our XCLASS and RADEX results is due to the LTE and non-LTE conditions assumed in each analysis. The caveat with the non-LTE calculation is that the estimated collision constants for all the transitions in this work do not exist for the kinetic temperatures estimated (higher than 50 K but lower than 100 K), and thus the RADEX results are extrapolated based on higher temperature conditions \citep[calculations from][]{Balanca2016SO2collisions}. Nevertheless, the result of this exercise is that both the kinetic temperature and column density at peak 2 can be higher than what we obtained with our LTE estimation, and that the volume density must be larger than $10^7$ cm$^{-3}$.


\section{Discussion\label{sec:discussion}}

\subsection{Origin of the \SOt peaks}

The two \SOt peaks can have different physical origins. As stated in Sect. \ref{sec:morphology}, peak 1 lies close to the base of the outflow, and the kinematic analysis of the \SOt brightest line shows that emission in this region follows the expected motion from an IRE. Peak 2, however, is not associated with CO emission from the outflow base (Fig. \ref{fig:SO2othertrans}) and its peak velocity is blueshifted with respect to the expected IRE motion. The size of the disk is too small \citep[radius $<30$ au,][]{Segura-Cox2018VANDAM} to influence the gas velocity at this position.

\begin{figure*}[htbp]
    \centering
    \includegraphics[width=0.33\textwidth]{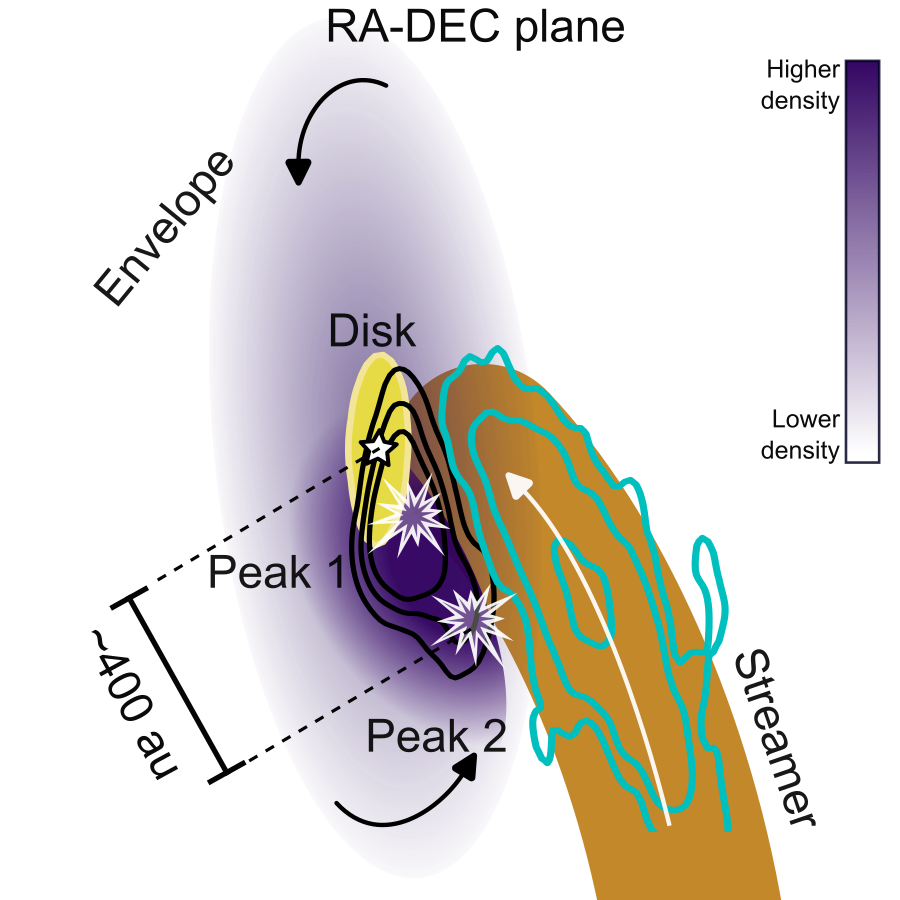}
    \includegraphics[width=0.33\textwidth]{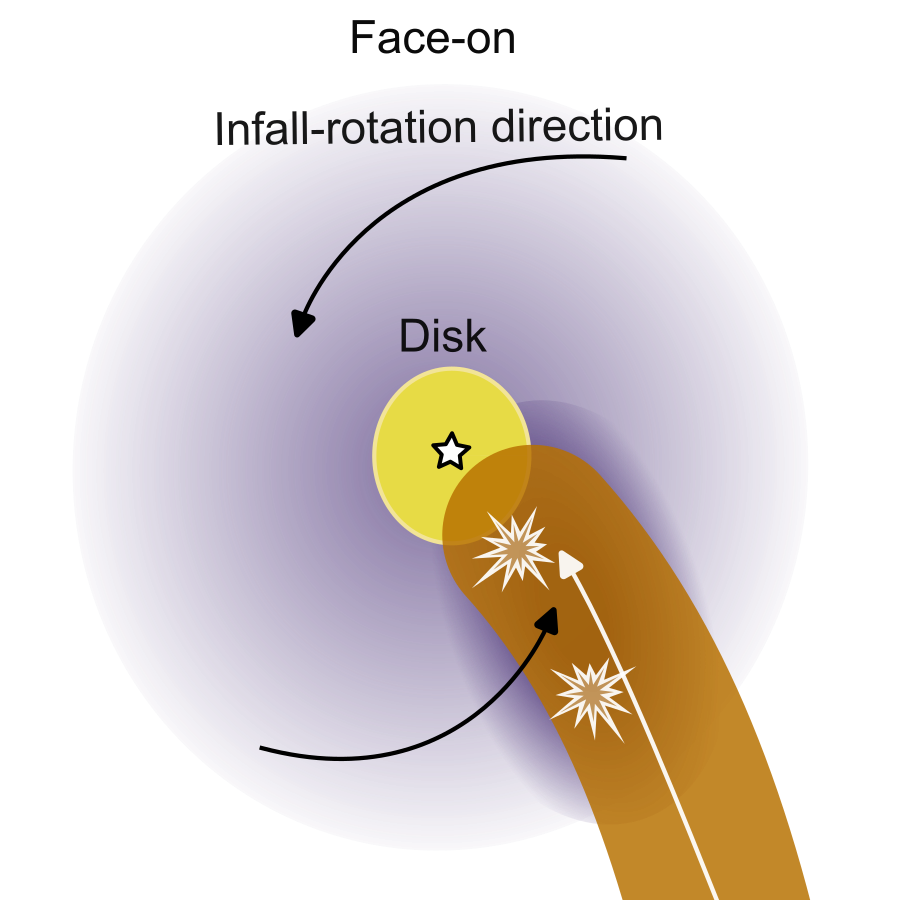}
    \includegraphics[width=0.33\textwidth]{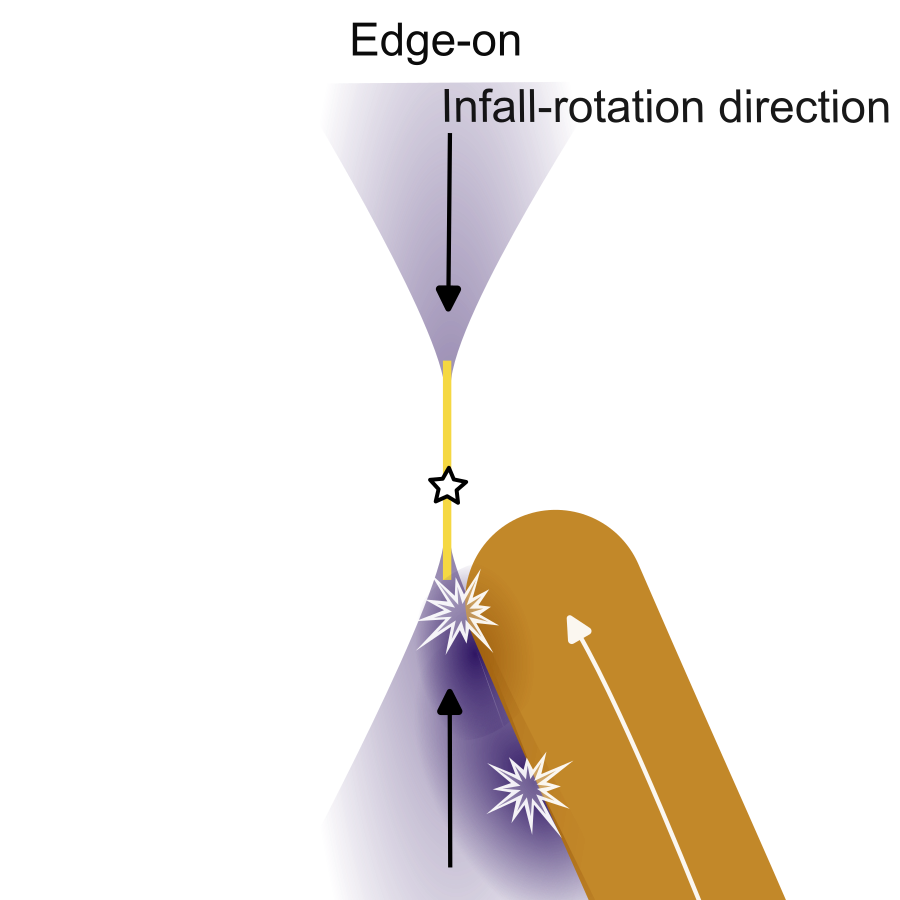}
    \caption{Schematic figure of the proposed interpretation of \SOt emission: the envelope (purple) impacts the streamer (brown) from the side as the latter falls toward the disk, causing a higher density and temperature and thus generating an \SOt peak ar around 400 au from the protostar (white star). Peak 1 is a shock, potentially in the disk-envelope interface, that is unclear if it is caused by the streamer or not. The peaks are shown with white explosion signs and labeled. The black arrows indicate the direction of envelope infall-rotation motions. The white arrow indicates the general motion of the streamer. Black and cyan contours show the brightness distribution of \SOt and \HtCO, respectively, from Fig. \ref{fig:mom0wh2co}. }
    \label{fig:summary}
\end{figure*}

We propose that peak 2 is caused by the shock between the inner envelope and the streamer as the latter moves toward the protostar, but before reaching the disk region of Per-emb 50. This is mostly supported by the shift of SO and \SOt emission toward blueshifted velocities close to the streamer. \HtCO emission tracing the streamer is accelerating toward blueshifted velocities, whereas the infall-rotation of the envelope, traced through both SO ad \SOt toward the south, is redshifted (Fig. \ref{fig:IREspectra}). 
The streamer and the \SOt peaks are not co-spatial because we are observing a flattened envelope from a close to edge-on configuration \citep[$i\sim70\deg$,][]{Segura-Cox2018VANDAM, Zhang2023PEACHES}, and the streamer hits one side of the inner envelope at an angle. We suggest that, as the streamer moves toward the disk, it pushes the envelope from one side, causing material to compresses within the envelope and heat up.
Thus, the relative motion between streamer and envelope causes a shock front, which we observe thanks to the SO and \SOt peak 2. If we were to observe the disk-envelope-system face on, \SOt and \HtCO emission would be co-spatial. This picture is visualized in Fig. \ref{fig:summary}. SO and \SOt have been associated to shocks produced by streamers in other sources \citep{Artur2022IRS44shocks, Garufi2022almadotshocks, Flores2023eDisk,Tanious2024L1489streamer}. Other shock tracing molecules such as \ce{SiO} (which is included in the PRODIGE setup) are not detected toward Per-emb 50, which suggests that the shock is not strong enough to sputter Si-material \citep[e.g.][]{Schilke1997sputter}. 
 This scenario of shocked compression is consistent with the observed polarization vectors toward peak 2 seen by \cite{Zhang2023PEACHES}, where they suggest that the different polarization fractions toward peak 2 with respect to the disk are due to a slow shock changing the properties of dust in that region.

\SOt peak 1 could also be due to the streamer, but given the presence of CO (Fig. \ref{fig:SO2othertrans}) and its proximity to the protostar (about 180 au), there are other possibilities that explain their presence. 
First, this could be a shock at the disk-envelope interface due to the infall from the envelope to the disk \citep[e.g.][]{Terebey2025infallouterdisk}.
This is supported by the fact that we see \SOt emission consistent with rotation both to the north and south of the protostar (Fig. \ref{fig:bayessigmasep}). This phenomenon has been used to explain the emergence of molecular emission close to the disk boundary in other protostars \citep{Sakai2014NatcentbarrierSO}. The offset, peak emission of SO and \SOt in this scenario can be explained by asymmetries in the envelope mass distribution; this asymmetry can be caused by the streamer, as suggested in \cite{Zhang2023PEACHES}, but can also be related to an asymmetric gas distribution in the envelope \citep[e.g.][]{Sakai2016centbarrierSOCS}. Recently, \cite{Liu2025SOSO2streamers} showed that SO and \SOt associated with streamers and envelope infall can be observed simultaneously. Second, it could be produced by a shock between the outflow and the envelope material at a distance of 100 to 200 au. This is supported by the coincidence between the \SOt peak 1, SO and CO emission following the outflow (Figs. \ref{fig:SO2othertrans} and \ref{fig:ap-spectraCO}). SO is known to trace outflows in young embedded sources \citep{Artur2023PEACHES}, and SO enhancements have been seen in outflow interactions with envelope material \citep{Oya2025SOinjetElias29}. A final possibility is that SO and \SOt near the protostar are tracing a disk wind. Both molecules have been used to detect and characterize disk winds in protostars \citep[e.g.][]{Tabone2017HH212SOdiskwind}, and the brightness distribution tracing the disk wind can be asymmetric \citep[][]{vantHoff2023eDiskoutflow, Mori2025simdiskwinds}. Higher resolution observations are needed to determine which scenario can explain the brightest \SOt peak.

\subsection{Shock conditions}

Assuming both peaks are generated by shocks, we make a rough estimate of the shock conditions based on the geometry of the system and the column densities of \SOt and SO. We obtained an estimate of the column density of SO ($N$(SO)) at the location of the \SOt peaks by fitting the observed SO $5_5-4_4$ beam averaged spectra at the location of both peaks using XCLASS. We use a similar procedure as in Sect. \ref{sec:so2lte-nonlte}, but we fixed the rotational temperatures to the ones found for \SOt (Table \ref{tab:LTEprops}), as $T_{\mathrm{rot}}$ cannot be constrained using only one SO transition. For peak 1, we obtain $N(\mathrm{SO}) \approx 4\times10^{15}$ cm$^{-2}$, and for peak 2,  $N(\mathrm{SO}) \approx 3\times10^{15}$ cm$^{-2}$. Comparing with the LTE column densities obtained for \SOt, the abundance ratio \SOt/SO is around 0.3 for peak 2 and about 0.2 for peak 1. Comparing with the shock models from \cite{vanGelder2021shockmodel} (their Fig. 10), the ratios roughly agree with a low velocity shock ($\sim2-3$ \kms), or intermediate velocity shock ($\sim5-7$ \kms) in the case the pre-shock volume density was lower (around 10$^6$ cm$^{-3}$) than the current (post-shock) density ($\sim10^7$ cm$^{-3}$, Sect. \ref{sec:so2lte-nonlte}). Given that our non-LTE values give a lower limit of around 10$^7$ cm$^{-3}$ for the current density (Appendix \ref{ap:nonLTE}), it is possible that the density of the gas increased locally thanks to the shock between envelope and streamer. According to this model, the temperature increase due to the shock could enhance SO and \SOt via their gas phase formation routes.

The derived velocities for the streamer and the envelope are consistent with the shock scenario suggested by the LTE conditions. To estimate the absolute velocity difference, we obtained the velocity vectors for the free-falling trajectory of the streamer, using the parameters from \cite{Valdivia-Mena2022prodigeI}, and the velocity vectors for the IRE model from \cite{Zhang2023PEACHES}. Then, we selected the IRE locations that intersected with the streamer model. The total, three-dimensional velocity difference between the modeled IRE and the modeled streamer at 400 au from the protostar is $\approx2.6$ \kms. This velocity difference is consistent within the expected velocity for the shock based on the abundance ratio described above. This is also the case the centrifugal radius of the streamer, at about 250 au: here, the absolute difference in velocity between both models is about 4 \kms. The higher velocity is consistent with a lower \SOt/SO abundance ratio under low velocity shocks \citep{vanGelder2021shockmodel}.

These are rough estimates of the potential shock conditions. Both the IRE from \cite{Zhang2023PEACHES} and the best free-fall model from \cite{Valdivia-Mena2022prodigeI} are estimations based on visual inspection instead of fitting, so the velocities obtained from them might not be the true velocities of the envelope and the streamer. Also, the comparison with \cite{vanGelder2021shockmodel} is an estimate, given our LTE approximation of the column densities. 
Nevertheless, our observations are not consistent with a stronger or faster shock. Given that SiO was not detected toward this source, as mentioned previously, the velocity of the streamer with respect to the envelope is not enough to sputter Si from the dust grains \citep{Schilke1997sputter, Caselli1997Jshock, Jimenez-Serra2008grains}. This low velocity is also not enough to sputter ice, therefore, any desorption in this region may be via thermal desorption due to the increase in temperature after the shock, in particular of more volatile species \citep{Aota2015SOshock}. This can affect the abundances of species with lower binding energies, such as CO and \ce{CH4}, and thus modify the abundance of this molecule in regions impacted by streamers. We note that streamers can produce stronger shocks in other regions, such as in IRS 63, where the abundance of \ce{D2CO} gas can be explained by the streamer shock as it sputters ices \citep{Podio2024streamer}. Further investigation into the impact locations of streamers toward disks and inner envelopes is needed to understand the range of chemical changes they can produce.

\section{Summary\label{sec:summary}}

In this work, we present NOEMA observations with the most extended configuration (A) toward the Class I protostar Per-emb 50. We analyzed three rotational transitions of \SOt wth $E_{\mathrm{up}}$ from 19 to 93 K, together with SO $5_{5} - 4_{4}$, \HtCO $3_{0,3} - 2_{0,2}$ and CO $2-1$ line emission. We find that \SOt is located offset from the protostar in two peaks toward the southwest, one at about 180 au (peak 1) and another at 400 au (peak 2). This paper focuses on analyzing the physical conditions based on \SOt and SO to explain their appearance. The physical interpretation of the molecular emission is summarized in Fig. \ref{fig:summary} and explained below.

\SOt emission traces part of an infalling-rotating envelope, traced by a narrow ($\sigma_v\sim0.7-0.8$ \kms) Gaussian component, as well as potential disk emission based on its central velocities, traced by a wider \SOt component ($\sigma_v\geq2$ \kms). Both peak 1 and 2 are part of the narrow Gaussian component. 
We estimated the physical conditions traced by \SOt using only the narrow components to separate potential disk emission from the peaks. 
An LTE analysis of the three \SOt transitions resulted in high column densities for \SOt (up to $10^{15}$ cm$^{-2}$ for the brightest peak) and temperatures of about 64 K for peak 1 and 46 K for peak 2. A non-LTE analysis corroborated the high column densities and showed that the volume density in peak 2 needs to be at least $10^{7}$ cm$^{-3}$ to explain the presence of the three transitions. 

Our analysis of \SOt and SO emission and abundance ratios suggest that peak 2, the farthest away from the protostar, is consistent with a slow shock ($\sim 3$ \kms) inside the inner envelope. Given the location of the peak (400 au away) and the presence of the streamer head toward the east of this peak, we propose that there is a shock due to the difference between the streamer and the envelope velocities toward the southwest. The disk-envelope-streamer system is edge-on with respect to the observer, thus we observe the shock from one side, where the envelope is compressed and heated due to the impact. 

\SOt peak 1 might also be caused by the streamer, but there are also other possibilities, such as shocks in the disk-envelope interface or interactions with the outflow. This peak is consistent with envelope motion (comparing to an infalling-rotating envelope model) and is potentially located at the outer part of the disk ($\sim 180$ au). The abundance ratios are also consistent with slow shocks in this region. Higher resolution data that resolves the gas disk are needed to investigate this peak's nature further. 

Our results indicate that streamers can affect the structure of both disks and envelopes: a streamer can affect the envelope as it comes into contact with the latter, causing a shocked and compressed layer where the two meet, and it can also impact the disk near its landing.
Our observations of Per-emb 50 show shocks due to a streamer and possibly also at the disk-envelope interface simultaneously, similar to observations in other embedded protostars \citep{Liu2025SOSO2streamers}.
Our results suggest that streamers can have a diversity of effects when infalling toward a young protostar, opening new avenues to explore the chemical inheritance from the ISM onto planet-forming disks.

\begin{acknowledgements}
We thank the anonymous referee for their constructive comments during peer review.
M.T.V-M. thanks Marta de Simone, Pooneh Nazari and Suchitra Narayanan for their insightful discussions about the contents of this article.
       M.T.V-M, J.E.P., C.G., P.C., Y.L.,  M.J.M, L.A.B,  D.S., Y-R.C., R. F. and K. S. 
acknowledge the support by the Max Planck Society. This project is co- funded by the European Union (ERC, SUL4LIFE, grant agreement No101096293). A.F. also thanks project PID2022-137980NB-I00 funded by the Spanish
Ministry of Science and Innovation/State Agency of Research MCIN/AEI/10.13039/501100011033 and by “ERDF A way of making Europe”.
D.~S. was funded by the Deutsche Forschungsgemeinschaft (DFG, German Research Foundation) – project number: 550639632. 
I.J-.S acknowledges funding from grant PID2022-136814NB-I00 funded by the Spanish Ministry of Science, Innovation and Universities/State Agency of Research MICIU/AEI/ 10.13039/501100011033 and by ``ERDF/EU”.
The authors thank the IRAM staff at the NOEMA observatory for their support in the observations and data calibration.
This work is based on observations carried out under projects number W22AH and L19MB with the IRAM NOEMA Interferometer. IRAM is supported by INSU/CNRS (France), MPG (Germany) and IGN (Spain).
      IRE models were calculated using FERIA \citep{Oya2022FERIA} available at \url{https://github.com/YokoOya/FERIA}.
      This research has made use of NASA's Astrophysics Data System Bibliographic Services.
\end{acknowledgements}

\bibliographystyle{aa} 
\bibliography{main} 

\begin{appendix}




\section{\SOt emission in comparison to other molecules\label{ap:zoomSO2}}

\begin{figure}[h]
    \centering
    \includegraphics[width=0.93\linewidth]{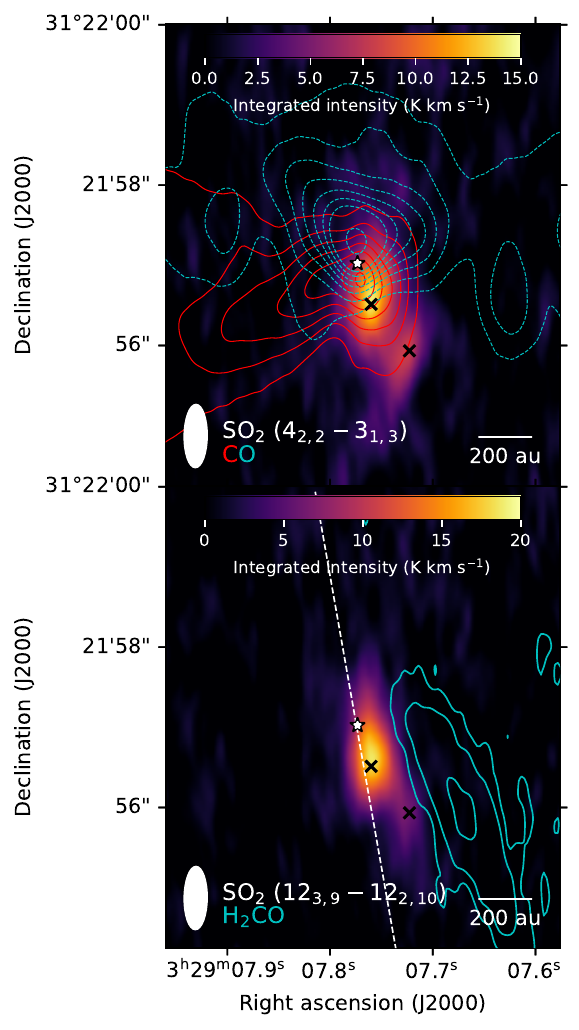}
    \caption{Integrated intensity of \SOt $4_{2,2}-3_{1,3}$ and $12_{3,9}-12_{2,10}$, compared with CO and \HtCO emission near the protostar. All \SOt transitions are integrated in the same velocity range as in Fig. \ref{fig:mom0wh2co}. Crosses mark the locations of peaks 1 and 2. Top: \SOt $4_{2,2}-3_{1,3}$ integrated intensity map. Red and blue contours mark the same CO contours as in Fig. \ref{fig:mom0wh2co}. Bottom: \SOt $12_{3,9}-12_{2,10}$ integrated intensity map. Blue contours represent the \HtCO integrated intensity, with contours drawn at 3, 5 and 10 times the rms of the map (0.7 K \kms). The dashed line represents the direction of the PV diagram in Fig. \ref{fig:SO2pvdiag}. }
    \label{fig:SO2othertrans}
\end{figure}

Figure \ref{fig:SO2othertrans} shows the integrated intensity of the \SOt transitions $4_{2,2}-3_{1,3}$ and $12_{3,9}-12_{2,10}$. Emission from these transitions is weaker than the $11_{1,11} - 10_{0,10}$ transition, but the distribution of the peaks is the same. We also plot CO contours and \HtCO contours over the \SOt transitions to compare their morphology. Redshifted CO emission coincides spatially with \SOt peak 1, although its maximum is closer to the protostar. \HtCO emission is not coincident with \SOt emission, with its maximum emission toward the southwest of the protostar.

We compared the emission from SO and \SOt with the expected emission from an infalling-rotating envelope (IRE). We first created a simulated data cube using FERIA \citep{Oya2022FERIA}, with the IRE parameters found by \cite{Zhang2023PEACHES} as input: a density power law $\alpha=-1.5$, envelope height scale $h(r)/r = 0.2$, an outer radius of 2.4\arcsec, a centrifugal barrier distance of 0.4\arcsec (119 au), projected centrifugal velocity $v_{\mathrm{rot, max}} \cos(i) = 3.5$ \kms and a disk inclination angle of $10\deg$ \citep[Table 4 of][]{Zhang2023PEACHES}. We then obtained a PV diagram of this simulated IRE using the same path as for SO and \SOt. The resulting PV diagrams of the simulated IRE are plotted in white contours in Fig. \ref{fig:SO2pvdiag}. We did not include the Keplerian rotating disk in the IRE model because we focused only on the bright, resolved \SOt emission. 
We note that the parameters from \cite{Zhang2023PEACHES} return a central mass of 0.85 \Msun, which differs from the mass of 1.7 \Msun determined in \cite{Valdivia-Mena2022prodigeI}. The centrifugal barrier radius $r_{\mathrm{cb}}$ and the rotation velocity at that location $v_{\mathrm{rot, max}}$ are related by the central mass by:
\begin{equation}
    r_{\mathrm{cb}} = \frac{2GM_{*}}{v_{\mathrm{rot, max}}^2},
\end{equation}
based on the equations of the IRE model in \cite{Oya2022FERIA}. After deprojecting the velocity at the centrifugal barrier by $\cos(i)$, the resulting mass from the IRE model parameters is 0.85 \Msun. The difference occurs because both masses are estimates based on adjusting models to the observations by visual inspection, and the observations in both cases do not resolve the disk. Therefore, the mass of Per-emb 50 is not well constrained.

\begin{figure}[htbp]
    \centering
    \includegraphics[width=0.45\textwidth]{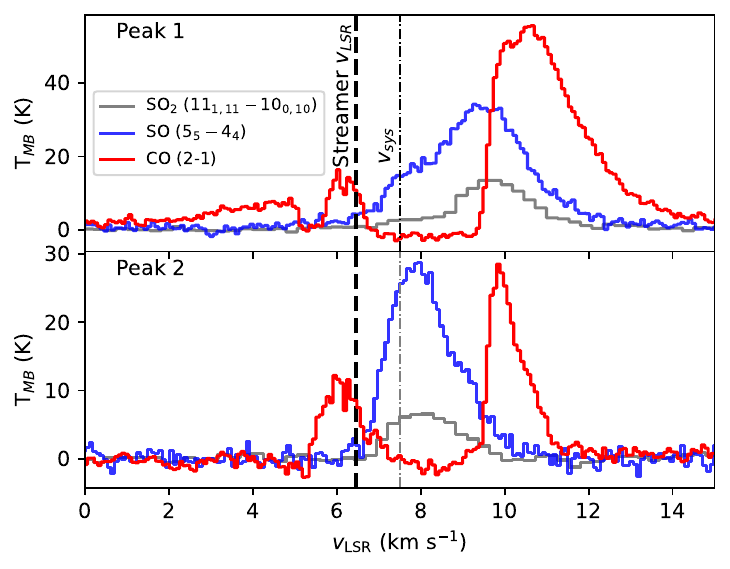}
    \caption{Same as Fig. \ref{fig:IREspectra} but with CO $2-1$ spectra at the location of the \SOt peaks. CO emission is drawn in solid red lines. The vertical dashed-dotted line marks the systemic velocity of the protostar (7.5 \kms), whereas the thick dashed line marks the velocity of the streamer at the same distance from the protostar (in radius) as peak 2 (6.45 \kms, Fig. \ref{fig:ap-H2CObestfit} middle. }
    \label{fig:ap-spectraCO}
\end{figure}

Figure \ref{fig:ap-spectraCO} shows the spectra of SO, \SOt and CO at the \SOt peak locations. CO has no emission observed between approximately 7 and 9.5 \kms due to large-scale cloud emission not recovered by our NOEMA observations. SO and \SOt emission in peak 1 have a velocity of 9.8 \kms, close to the apparent peak of CO at that location (at 10.4 \kms), and so it is possible that SO and \SOt emission are associated to the outflow at that location. However, CO emission in peak 2 is dominated by large-scale emission and is farther away from the apparent redshifted outflow lobe base (Fig. \ref{fig:SO2othertrans} top).  

\section{Streamer kinematics\label{ap:streamerH2CO}}

\begin{figure}[!ht]
    \centering
    \includegraphics[width=0.9\linewidth]{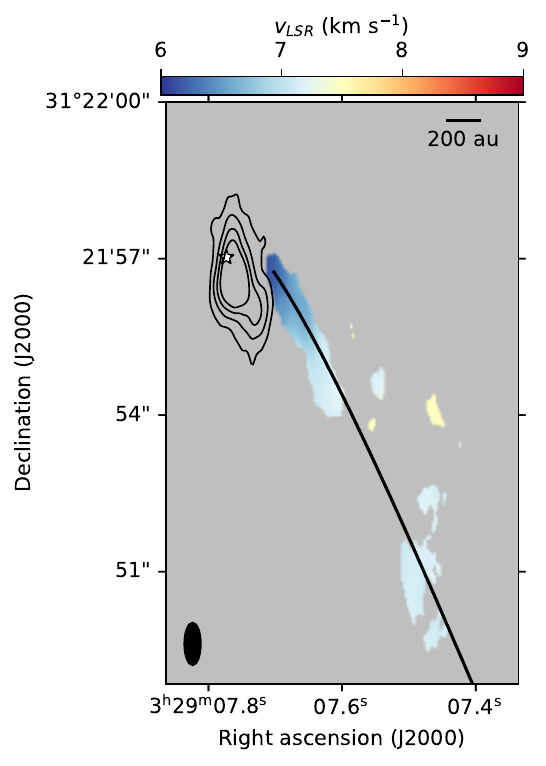}
    \caption{Best fit central velocity for the Gaussian fit to \HtCO $3_{0,3} - 2_{0,2}$, for emission with $SNR>5$. Black contours mark 5, 10, 15 and 20 times the rms level (0.8 K \kms) of the \SOt $11_{1,11} - 10_{0,10}$ integrated intensity image from Fig. \ref{fig:mom0wh2co}. The black curve represents the streamer trajectory from \cite{Valdivia-Mena2022prodigeI}. An ellipse in the bottom left corner represents the beam. The scalebar in the top right shows a length of 200 au.}
    \label{fig:ap-H2CObestfit}
\end{figure}

The \HtCO spectra across the whole length of the streamer show one clear Gaussian peak, consistent with previous results from \cite{Valdivia-Mena2022prodigeI}. We fit one Gaussian to each spectrum with $SNR>5$ using \texttt{pyspeckit} \citep{pyspeckit2011,pyspeckit2022}. The best fit central velocities $v$ for the new NOEMA map are shown in Fig. \ref{fig:ap-H2CObestfit}. The velocities decrease down to approximately 6.2 \kms at the tip of the streamer. However, at the projected distance of \SOt peak 2 (400 au), the average velocity in a beam of \HtCO emission is 6.45 \kms. We note that there is no \HtCO emission at the location of \SOt peaks.

\section{Nested sampling fits of \ce{SO2} \label{ap:gaussfits}}

\begin{figure}[!ht]
    \centering
    \includegraphics[width=0.95\linewidth]{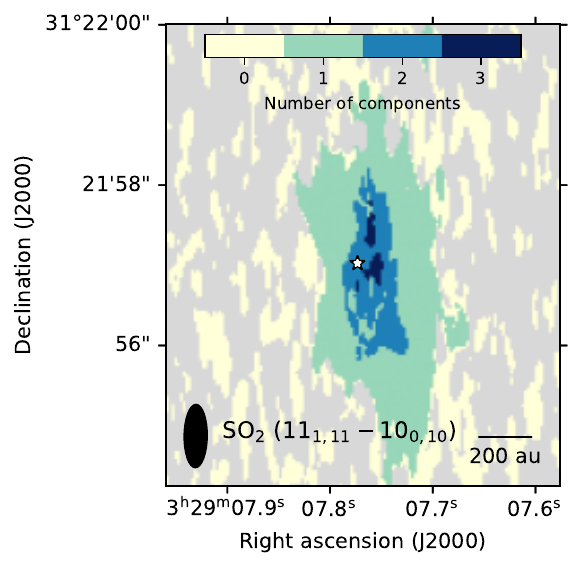}
    \caption{Number of Gaussian components that best fit the \SOt $11_{1,11} - 10_{0,10}$ spectra. The white star marks the position of the protostar. The black ellipse in the bottom left corner shows the size of the beam. A scalebar represents a physical size of 200 au.}
    \label{fig:npeaks-so2}
\end{figure}

We fit up to three Gaussian curves to each of the spectra in the brightest \SOt transition ($11_{1,11} - 10_{0,10}$) which has $SNR > 5$. We applied Bayesian model selection through the Nested Sampling algorithm to fit and select the number of components that at the same time. A full description of this line decomposition method is found in \cite{Sokolov2020BayesFit}. We used the \texttt{pymultinest} Python package \citep{Buchner2014X-raypymultinest} to run the Nested Sampling algorithm in Python. We used \texttt{pyspecnest}  \citep{Sokolov2020BayesFit} to introduce \texttt{pyspeckit} models to the \texttt{pymultinest} code. We selected the best model by using the standard threshold of $\ln(K_{n-1}^{n})= 5$, where $n$ is the number of Gaussian components of the model being compared against a model with $n-1$ components. In this case, $n=0$ corresponds to background noise. Figure \ref{fig:npeaks-so2} shows the number of Gaussian components that best fit each pixel of the \SOt $11_{1,11} - 10_{0,10}$ data cube. 
Close to the protostar, there are two Gaussian components recognized through Nested Sampling, and a few pixels toward the west of the continuum peak are best fitted with three Gaussians. 

\begin{figure}[!ht]
    \centering
    \includegraphics[width=0.95\linewidth]{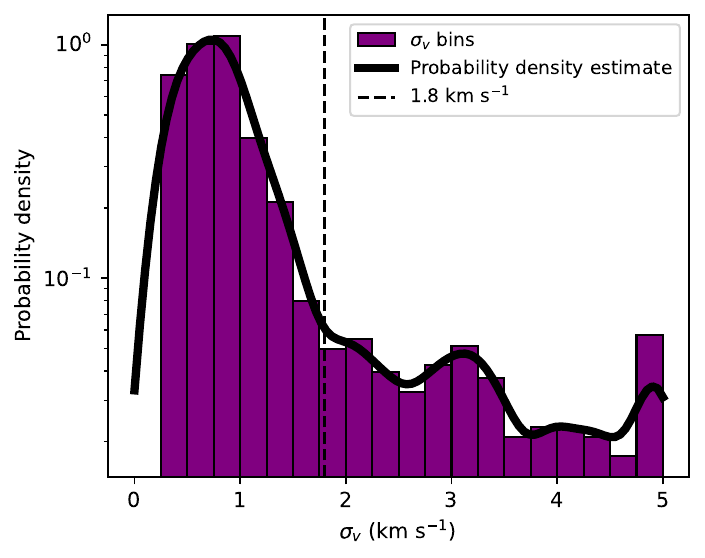}
    \caption{Histogram and probability density estimate of the $\sigma_v$ values obtained for all Gaussian fits to the \SOt  line cube. The histogram is normalized by density. The dashed line marks the position of the first inflection point of the PDE (1.8 \kms). }
    \label{fig:sigmasepSO2}
\end{figure}

Figure \ref{fig:sigmasepSO2} shows the distribution of $\sigma_v$ values from all Gaussian components in the map, together with the estimated probability density function (PDF) of $\sigma_v$. We first made a histogram of the obtained $\sigma_v$ values for all Gaussian fits found in the \SOt map. Then, we estimated the PDF of the $\sigma_v$ distribution by calculating its Kernel Density Estimate (KDE), using the scipy.stats function gaussian\_kde. The majority Gaussian components have $\sigma_v<1$, with a peak around 0.7 \kms, but there are also additional peaks of $\sigma_v$, one at around 2.2 \kms and a third around 3.1 \kms. We located the inflection point between the first two peaks at $\sigma_v=1.8$ \kms.


\section{Non-LTE calculations\label{ap:nonLTE}}

We calculated line ratios between the different transitions of \SOt in grids of kinetic temperature and \SOt column density, using the non-LTE radiative transfer code RADEX \citep{vanderTak200RADEX}. For $n_{\ce{H2}} = 10^6$ cm$^{-3}$, RADEX was run with a set of 50 equally spaced temperatures from 30 to 100 K and 50 equally spaced column density values from $5\times10^{12}$ to $5\times10^{15}$ cm$^{-2}$. For $n_{\ce{H2}} = 10^7$ and $10^8$ cm$^{-3}$, the grid was run with 50 equally spaced temperatures from 20 to 80 K and 50 equally spaced column density values from $10^{14}$ to $5\times10^{16}$ cm$^{-2}$. Combinations with column density higher than $5\times10^{15}$ cm$^{-2}$ for $n_{\ce{H2}} = 10^6$ cm$^{-3}$ did not converge. Besides these values, we used the default values for the radiation model (Uniform sphere) and background temperature (2.73 K). 

\begin{figure}
    \centering
    \includegraphics[width=0.3\textwidth]{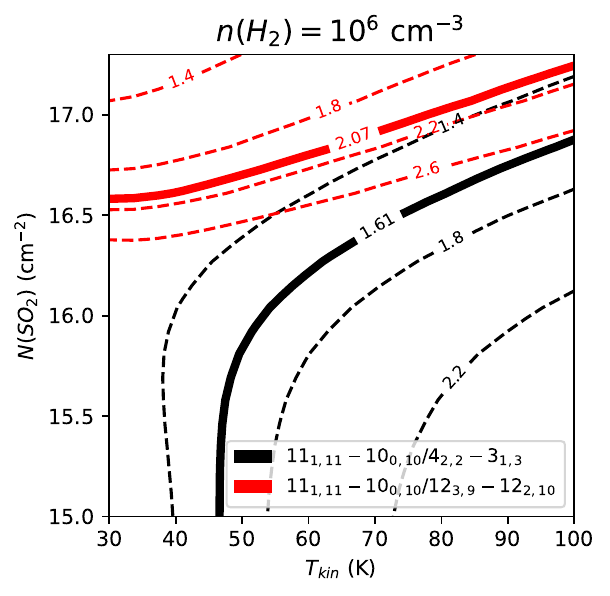}
    \includegraphics[width=0.3\textwidth]{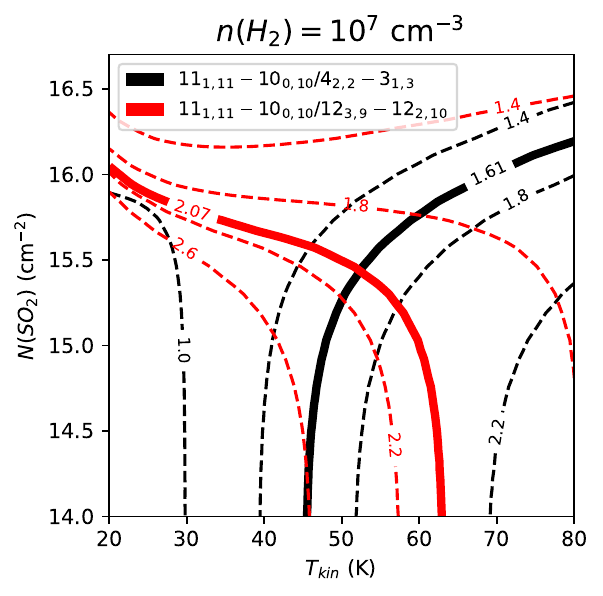}
    \includegraphics[width=0.3\textwidth]{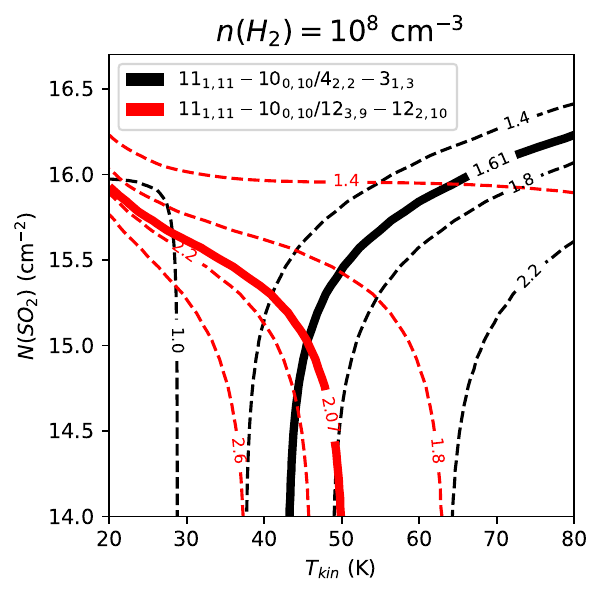}
    \caption{Line ratios between \SOt $11_{1,11}-10_{0,10}$, $4_{2,2}-3_{1,3}$ and $12_{3,9}-12_{2,10}$ transitions resulting from different combinations of $T_{\mathrm{kin}}-N(\mathrm{SO}_2)$ obtained with RADEX. Black lines show the resulting $11_{1,11}-10_{0,10}$ over $4_{2,2}-3_{1,3}$ line ratio, whereas red lines show the $11_{1,11}-10_{0,10}$ over $12_{3,9}-12_{2,10}$ line ratio. The thick black and red lines mark the line ratios found in the spectra of peak 2. The dashed black and red lines indicate where other line ratio values are located in the grid. Top: results for $n_{\mathrm{H}_2} = 10^6$ cm$^{-3}$. Middle: results for $n_{\mathrm{H}_2} = 10^7$ cm$^{-3}$. Bottom: results for $n_{\mathrm{H}_2} = 10^8$ cm$^{-3}$. }
    \label{fig:radexresults}
\end{figure}

We recovered the line intensities from the RADEX outputs for each combination of parameters. It was not possible to replicate the line intensities for the three transitions assuming a beam filling fraction of 1. To avoid assuming a beam filling factor for emission of an unknown size, we instead used the line ratios of the different transitions to comper with our observed \SOt emission. We plotted where the line ratios obtained from our \SOt observations of peak 2 lie in the RADEX grids. The resulting positions in the $T_{\mathrm{kin}}-N(\ce{SO2})$ grids for each $n_{\ce{H2}}$ are shown in Fig. \ref{fig:radexresults}.

Our RADEX calculations show that there is no $T_{\mathrm{kin}}-N(\ce{SO2})$ possible combination for $n_{\ce{H2}} = 10^6$ cm$^{-3}$ where the line ratios can coexist. For $n_{\ce{H2}} = 10^7$ and $10^8$ cm$^{-3}$, there exist an intersection area. The line ratios intersect at $T_{\mathrm{kin}}=51$ K and log $N(\ce{SO2}) = 15.4$ cm$^{-2}$ for $n_{\ce{H2}} = 10^7$ cm$^{-3}$, and at $T_{\mathrm{kin}}=45$ K and $\log N(\ce{SO2}) = 15.0$ cm$^{-2}$ for $n_{\ce{H2}} = 10^8$ cm$^{-3}$, so the higher $n_{\mathrm{H}_2}$, the lower the required $T_{\mathrm{kin}}-N(\ce{SO2})$ to replicate the observed line ratios.



\end{appendix}
\end{document}